\documentclass[twocolumn,showpacs,twoside,10pt,aps,prl,superscriptaddress,longbibliography]{revtex4-2}
\usepackage{graphicx}
\usepackage{epsfig}
\usepackage{epsf}
\usepackage{amsmath,amsfonts,amssymb,natbib}
\usepackage{url,textcase,bm}
\usepackage{amsthm}
\usepackage{multirow}
\usepackage{mathrsfs}
\usepackage[colorlinks=true,citecolor=blue,urlcolor=blue]{hyperref}
\newcommand\rket[1]{|#1\rangle}

\begin{document}

\title{
Coherence properties of rare-earth spins in micrometer-thin films 
}

\author
{
Zihua Chai$^{1,2\ast}$, 
Zhaocong Wang$^{3\ast}$, 
Xinghang Chen$^{1,2}$,
Quanshen Shen$^{1,2}$, 
Zeyu Gao$^{1,2}$, 
Junyu Guan$^{1,2}$, \\
Hanyu Zhang$^{1,2}$, 
Ya Wang$^{1,2,4}$,
Yang Tan$^{3\dag}$, 
Feng Chen$^{3}$, 
Kangwei Xia$^{1,2,4\dag}$
\\
\normalsize{$^{1}$ CAS Key Laboratory of Microscale Magnetic Resonance and School of Physical Sciences,}\\
\normalsize{University of Science and Technology of China, Hefei 230026, China.}\\
\normalsize{$^{2}$ Anhui Province Key Laboratory of Scientific Instrument Development and Application,}\\
\normalsize{University of Science and Technology of China, Hefei 230026, China.}\\
\normalsize{$^{3}$ School of Physics, Shandong University, Shandong, Jinan 250100, China.}\\
\normalsize{$^{4}$Hefei National Laboratory, University of Science and Technology of China, Hefei 230088, China.}\\
\normalsize{$^{\ast}$ These authors contributed equally to this work.}\\
\normalsize{$^\dag$ E-mail: tanyang@sdu.edu.cn, kangweixia@ustc.edu.cn}
}

\begin{abstract}
Rare-earth ions in bulk crystals are excellent solid-state quantum systems in quantum information science, owing to the exceptional optical and spin coherence properties. However, the weak fluorescence of single rare-earth ions present a significant challenge for scalability, necessitating the integration into micro-cavities. Thin films serve as a promising material platform for the integration, yet the fabrication without compromising the properties of the materials and rare-earth ions remains challenging. In this work, we fabricate micrometer-thin yttrium aluminum garnet (YAG) films from bulk crystals using ion implantation techniques. The resulting films preserve the single-crystalline structure of the original bulk crystal. Notably, the embedded rare-earth ions are photo-stable and exhibit bulk-like spin coherence properties. Our results demonstrate the compatibility of bulk-like spin properties with the thin-film fabrication technique, facilitating the efficient integration of rare-earth ions into on-chip photonic devices and advancing the applications of rare-earth ions systems in quantum technologies.
\end{abstract}

\maketitle

\section{introduction}
Rare-earth ions in bulk crystals are emerging platforms in quantum information science. Due to the excellent optical and spin coherence properties, rare-earth ions have been utilized in the field of quantum memories and quantum network. Landmark experimental demonstration includes a six-hour coherence time and one-hour quantum memories \cite{zhong_optically_2015, ma_onehour_2021}, spectral multiplexing \cite{afzelius_multimode_2009, saglamyurek_broadband_2011, sinclair_spectral_2014}, and entanglement distribution with quantum memories \cite{liu_heralded_2021, lago-rivera_telecom_heralded_2021}, et al. Detecting and manipulating single rare-earth ion can further utilize the properties and enhance the scalability, which is recently demonstrated with various platforms \cite{kolesov_optical_2012, kolesov_mapping_2013, siyushev_coherent_2014, zhong_optically_2018, dibos_atomic_2018, chen_parallel_2020, kindem_control_2020, ruskuc_nuclear_2022}. Due to the weak transition strength between $\rm 4f$ states, optical detection of single rare-earth ions is challenging. Effective integration of rare-earth ions and optical cavities is essential to enhance the light-matter interaction and facilitate connections between different nodes with photons, composing the key component of quantum networks \cite{reiserer_colloquium_2022, ruskuc_scalable_2024}. 

Cavity enhancement requires strong confinement of the light field, which necessitates the use of micro-cavities and imposes constraints on the dimensions of the host material \cite{kinos2021roadmaprareearthquantumcomputing, becher_2023_2023}. Thin films present a scalable material platform for integrating rare-earth ions into micro-cavities while allowing intrinsic optical and spin coherence properties in bulk materials to be preserved.
The integration, for example with Fabry-P{\'e}rot cavities, further necessitates an atomically flat surface to minimize scattering loss and a micrometer-level thickness to reduce the optical mode volume and enhance the Purcell effect of cavities \cite{meng_solid-state_2024}. Various methods have been proposed and experimentally demonstrated to obtain thin films \cite{zhong_emerging_2019,scarafagio_ultrathin_2019, balasa_rare_nodate,merkel_coherent_2020, ulanowski_spectral_2022, ulanowski_spectral_2024}. In the case of certain materials, such as $\rm Y_2 O_3$, thin films can be chemically synthesized with bulk-like optical properties \cite{scarafagio_ultrathin_2019, balasa_rare_nodate}. The extension to other materials and rare-earth ion platforms remains challenging due to the diverse material properties. 
Top-down methods, such as reactive-ion etching and chemo-mechanical polishing, can fabricate films from bulk crystals down to thicknesses on the order of tens of micrometers, while preserving the intrinsic properties \cite{merkel_coherent_2020, ulanowski_spectral_2022, ulanowski_spectral_2024}. However, achieving further thinning is difficult due to the high costs and failure rates. The limited thickness and configuration of these films constrain the size of cavities, limiting the mode volume and the performance. Fabricating transferable nanometer/micrometer-thin films is still challenging in the study of rare-earth ions, and the coherence properties of rare-earth ions in the thin films warrant exploration.

\begin{figure*}[t]
\includegraphics[width=1.6\columnwidth]{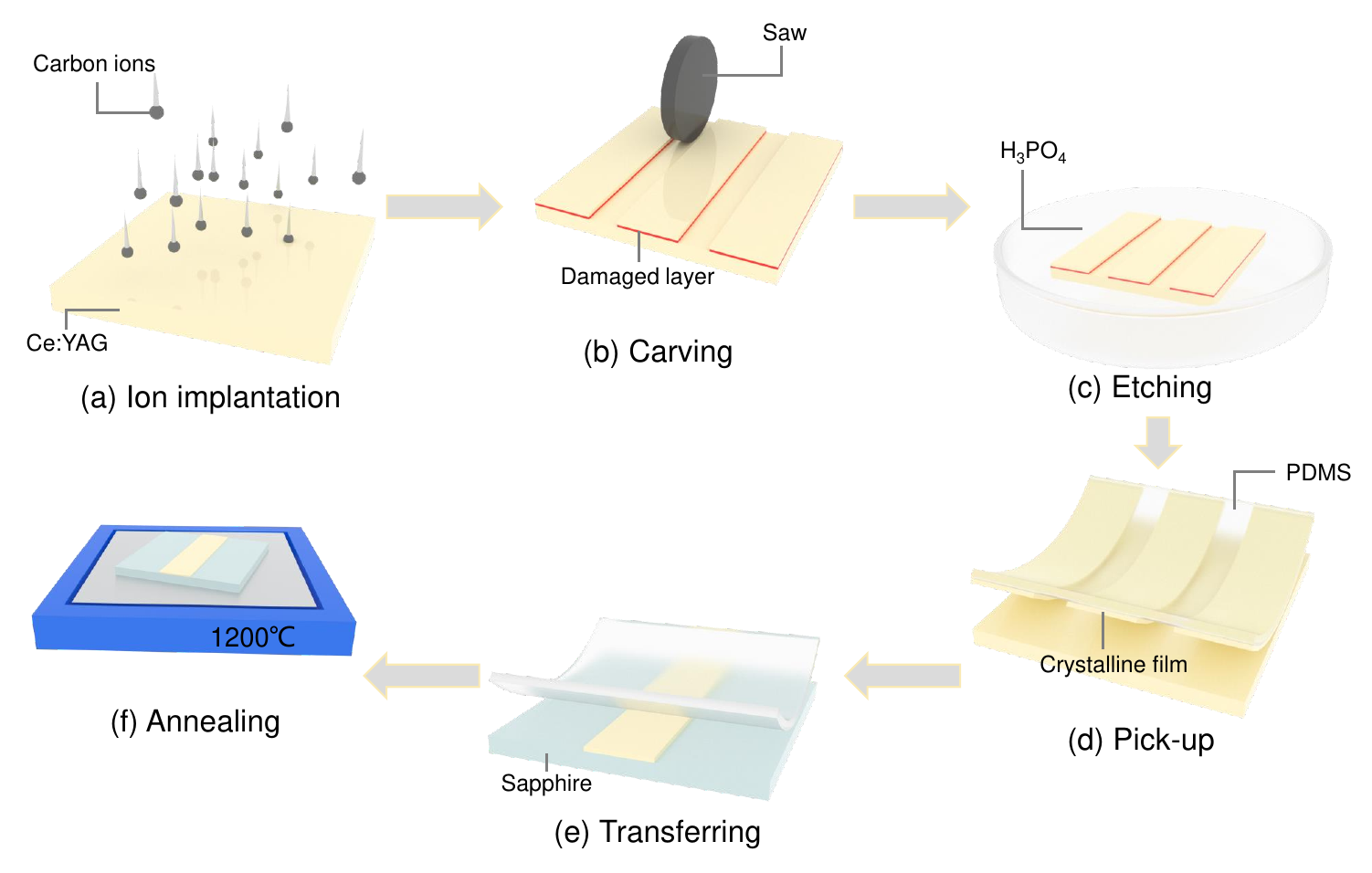}
\caption{
Preparation process of the YAG film sample. }
\label{fig1}
\end{figure*}


The "smart-cut" method has been applied to various materials to fabricate tunable single-crystalline thin films for integrated photonic devices, including diamond \cite{piracha_scalable_2016, guo_tunable_2021}, lithium niobate \cite{dutta_integrated_2020,wang_mathrmermathrmlimathrmnbmathrmo_3_2022, yang_controlling_2023} and yttrium aluminum garnet (YAG) \cite{li_optically_2023,li_27_2023}, exhibiting excellent material properties. The high-energy and high-dose ion implantation used in the fabrication processes introduce defects in the crystal lattice, potentially affecting the optical and spin properties of the embedded color centers. Consequently, the investigation of these properties is crucial for the development of nanophotonic quantum devices.
While bulk-like optical properties of rare-earth ions have been demonstrated in lithium niobate thin films \cite{dutta_integrated_2020,wang_mathrmermathrmlimathrmnbmathrmo_3_2022, yang_controlling_2023}, the strong nuclear spin bath and complex substitutional sites present significant challenges for spin control. As a result, the spin properties of rare-earth ions in "smart-cut" thin films are still not investigated. In contrast, yttrium-based materials, with stable and predictable substitutional sites, have been utilized to demonstrate quantum control over rare-earth spins \cite{siyushev_coherent_2014, chen_parallel_2020, kindem_control_2020, ruskuc_nuclear_2022, ruskuc_scalable_2024}. Thin films of yttrium-based crystals fabricated by "smart-cut" method could offer a promising integration of advancements in spin control and cavity-enhancement technologies. 
In this work, we investigate the spin coherence properties of rare-earth ions embedded in "smart-cut" YAG thin films. Our findings indicate that crystalline structure of the original crystal is well-preserved, and the embedded rare-earth spins exhibit coherence properties comparable to those in bulk crystals. These findings demonstrate that the "smart-cut" technique enables the fabrication of thin films while maintaining bulk-like spin coherence properties, highlighting the potential to develop nanophotonic quantum devices utilizing rare-earth ions.

\section{results}
\begin{figure}
\includegraphics[width=1\columnwidth]{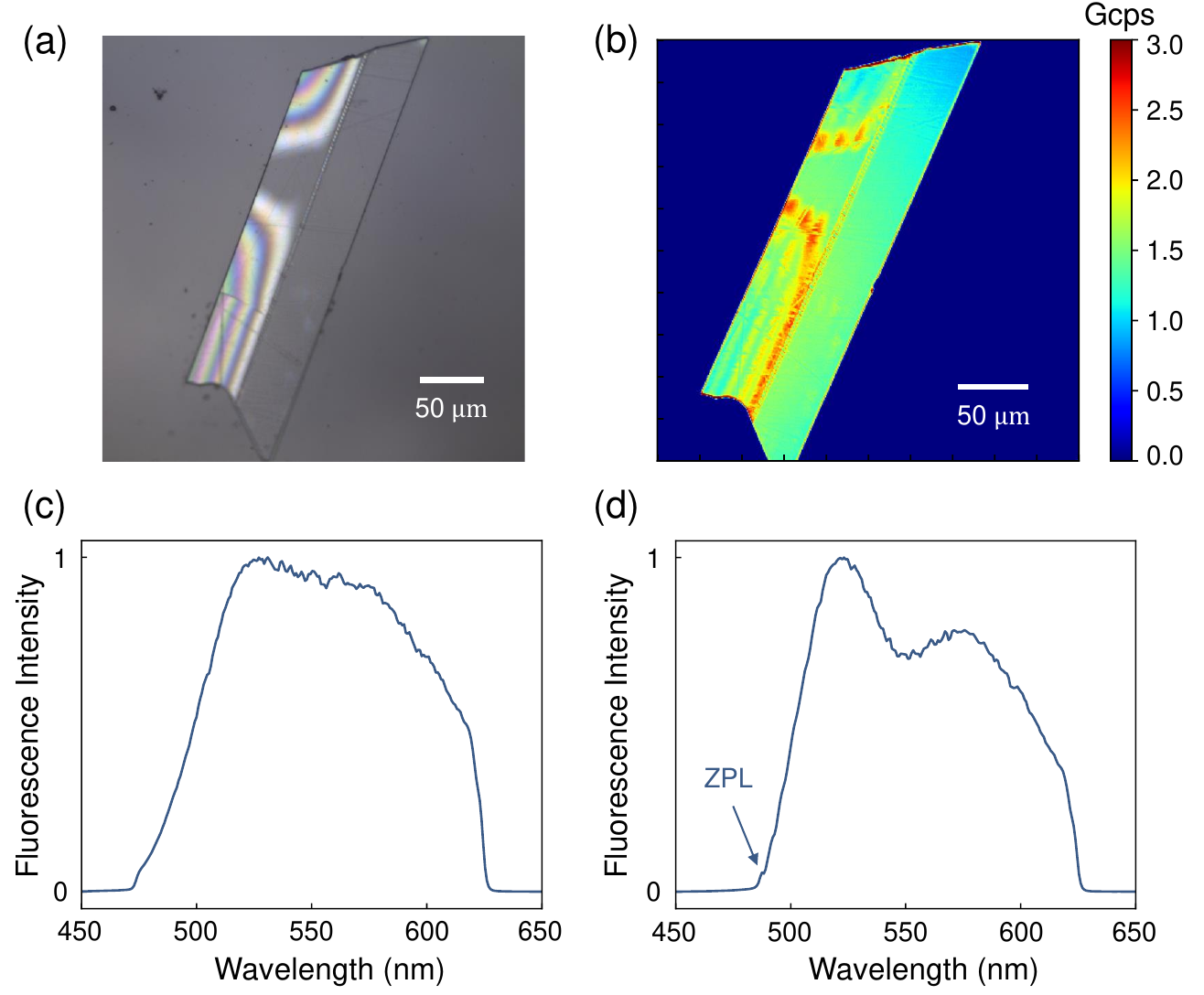}
\caption{
General properties of YAG films. 
(a) The microscope image of the transferred YAG film on sapphire substrate. 
(b) Photoluminescence image of the film sample with the excitation of femtosecond laser at 450 nm. 
(c) Emission spectrum of cerium ions at room temperature. 
(d) Emission spectrum of cerium ions at cryogentic temperature. 
}\label{fig2}
\end{figure}

The fabrication processes of the YAG film samples are illustrated in Fig.~\ref{fig1}. A Ce:YAG bulk crystal with dimensions of $5 \times 5\times 0.5\ \rm{mm^3}$ is used as the source crystal. The crystal is [110]-oriented and the concentration of cerium ions is 0.1\%. Carbon ($\rm C^{3+}$) ions are implanted into the YAG crystal at an energy of $\rm 6\ MeV$ and a dose of $\rm 2 \times 10^{15}\ ions\ cm^{-2}$ (Fig.~\ref{fig1}(a)),  creating a damaged layer beneath the surface. The incident direction of the carbon beam is at an angle of 7° off the normal direction of the facet to avoid the channeling effect. Following ion implantation, grooves are created on the surface of the bulk using a wafer saw (DISCO Co., P1A851-SD4000-R10-B01). The width and depth of the grooves are set as $\rm 300\  \mu m$ and $\rm 5\ \mu m$ to fully expose the damaged layer (Fig.~\ref{fig1}(b)). The sample is then immersed in a solution of 80\% phosphoric acid and heated to $\rm 120\ ^{\circ} C$ for 8 hours (Fig.~\ref{fig1}(c)) to exfoliate the YAG films. The films are collected and picked up with Polydimethylsiloxane (PDMS), as shown in Fig.~\ref{fig1}(d). Subsequently, the films are transferred and bonded to sapphire substrates using a PDMS-assisted site-specific transfer method (Fig.~\ref{fig1}(e)). Finally, the bonded film sample is annealed at 1200 $\rm ^{\circ} C$ for an hour to annihilate the crystal damages (Fig.~\ref{fig1}(f)). Fig.~\ref{fig2}(a) shows the microscope image of a film sample. Due to the protrusions on the sapphire substrate, the film sample is not bonded well. The width of the film exceeds 80 $\mu {\rm m}$, which is sufficient for the applications of nanophotonics. The characteristic thickness of the films is 1.22 $\rm{\mu m}$, and the surface roughness is measured to be 0.717 nm rms. Addtionally, the characteristic height variation over a scanning range of 85 $\mu {\rm m}$ is 14.4 nm \cite{SM}. 
The above properties make the films suitable for the integration of micro-cavities, demonstrating that the fabrication method is clear and scalable for producing transferrable YAG films for rare-earth ions experimental research. 

\begin{figure*}[ht]
\includegraphics[width=1.6\columnwidth]{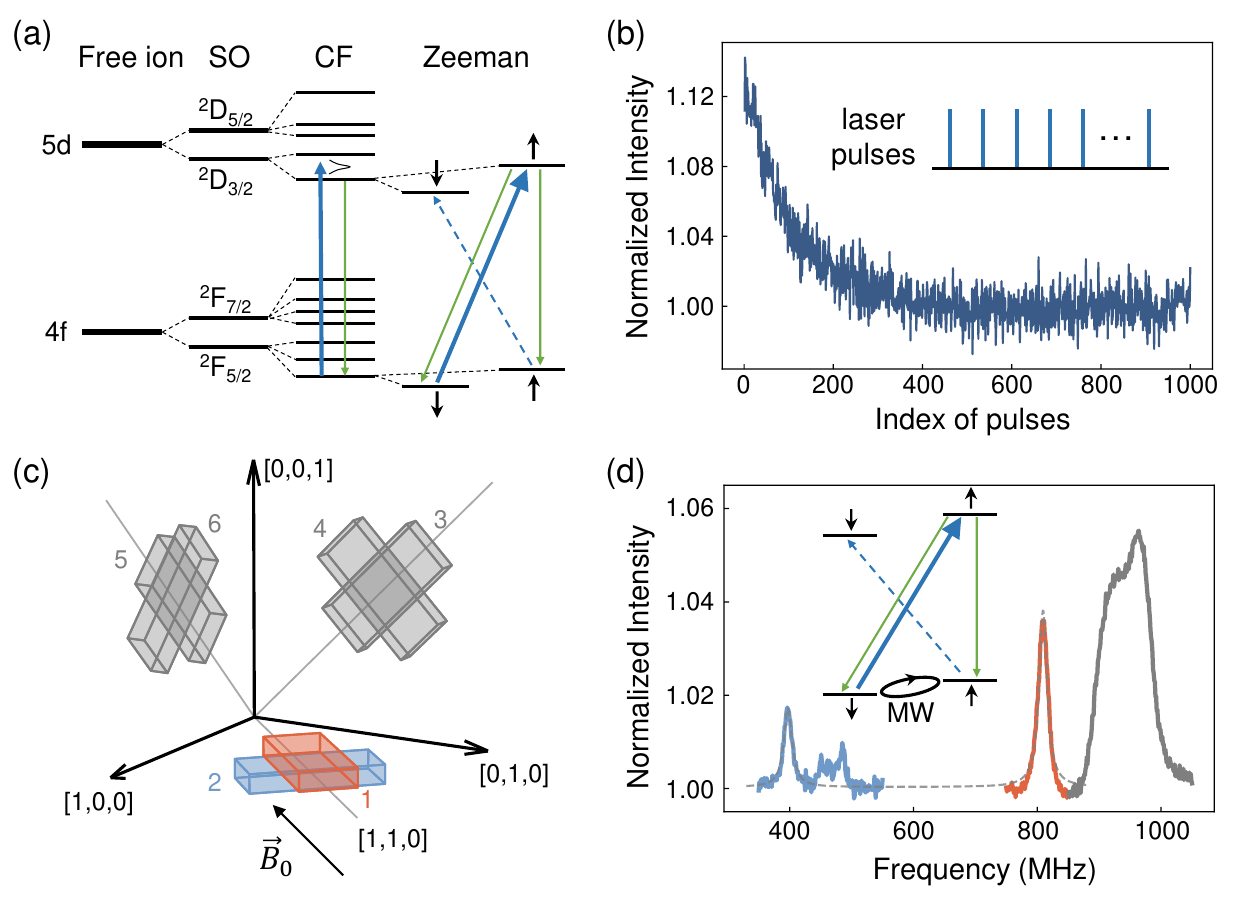}
\caption{The initialization and magnetic resonance of cerium electron spins at cryogenic temperature.
(a) Energy level structure of cerium ions in YAG crystal. The blue arrows represent laser excitation using circularly polarized laser. The green arrows represent the fluorescence process.
(b) The spin initialization dynamics of cerium ions in YAG film. The inset shows the laser pulses sequence used in the experiment. The fluorescence photons after each laser pulse are accumulated and normalized.
(c) The six orientations of cerium ions in YAG lattice. The magnetic field is applied parallel to [110] direction. The frame axes represent the crystallographic orientations. The dimensions of the cuboids represent the principal axis of the g-tensor, where the longest (intermediate/shortest) dimension corresponds to $g_x\ (g_y/g_z)$, respectively.
(d) Continuous-wave ODMR spectrum of cerium ions ensemble in YAG film. The blue (orange/gray) spectrum corresponds to the orientations of the same color in (c). 
}\label{fig3}
\end{figure*}

Based on the films, we perform experiments on the embedded rare-earth ions to investigate the spin properties. Among the various rare-earth ions, cerium ions stand out due to the ability to be excited with a laser at $\rm 450\ nm$, which matches the phonon-assisted absorption sideband of $\rm 4f-5d$ transition (Fig.~\ref{fig3}(a)), and efficiently read out through the $\rm 5d-4f$ fluorescence, even in the absence of cavity-enhancement. Additionally, cerium spins, which resides in the 4f orbital, exhibit Zeeman splitting that can be tuned to match the energy splitting of other rare-earth ions, making them suitable for probing the noise environment affecting rare-earth spins. A home-built optically detected magnetic resonance (ODMR) setup is utilized to locate the film, excite the cerium ions, and probe the properties of cerium electron spins. The confocal microscope in the setup is similar to the one used in prior studies \cite{kolesov_mapping_2013,siyushev_coherent_2014}. A femtosecond Ti:Sapphire laser is frequency-doubled to 450 nm with a second harmonic generation crystal to excite the cerium ions, and the repetition rate of the laser pulses is controlled by a pulse picker. The resulted fluorescence emitted from the $\rm 5d-4f$ transition is detected within a wavelength range of $\rm 475-625\ nm$ window with an avalanche photodiode. To prevent potential damage from high fluorescence intensity, an OD3 neutral density filter is incorporated for pre-attenuation \cite{SM}. Microwave control is realized through an arbitrary wave generator (AWG), while a pulse generator controls the timing sequences of laser and microwave with high time-resolution digital pulses. Due to the fast spin-lattice relaxation of cerium spins at room temperature, a cryostat is utilized to cool the film sample down to $\rm 6\ K$. 

\begin{figure*}
\includegraphics[width=2\columnwidth]{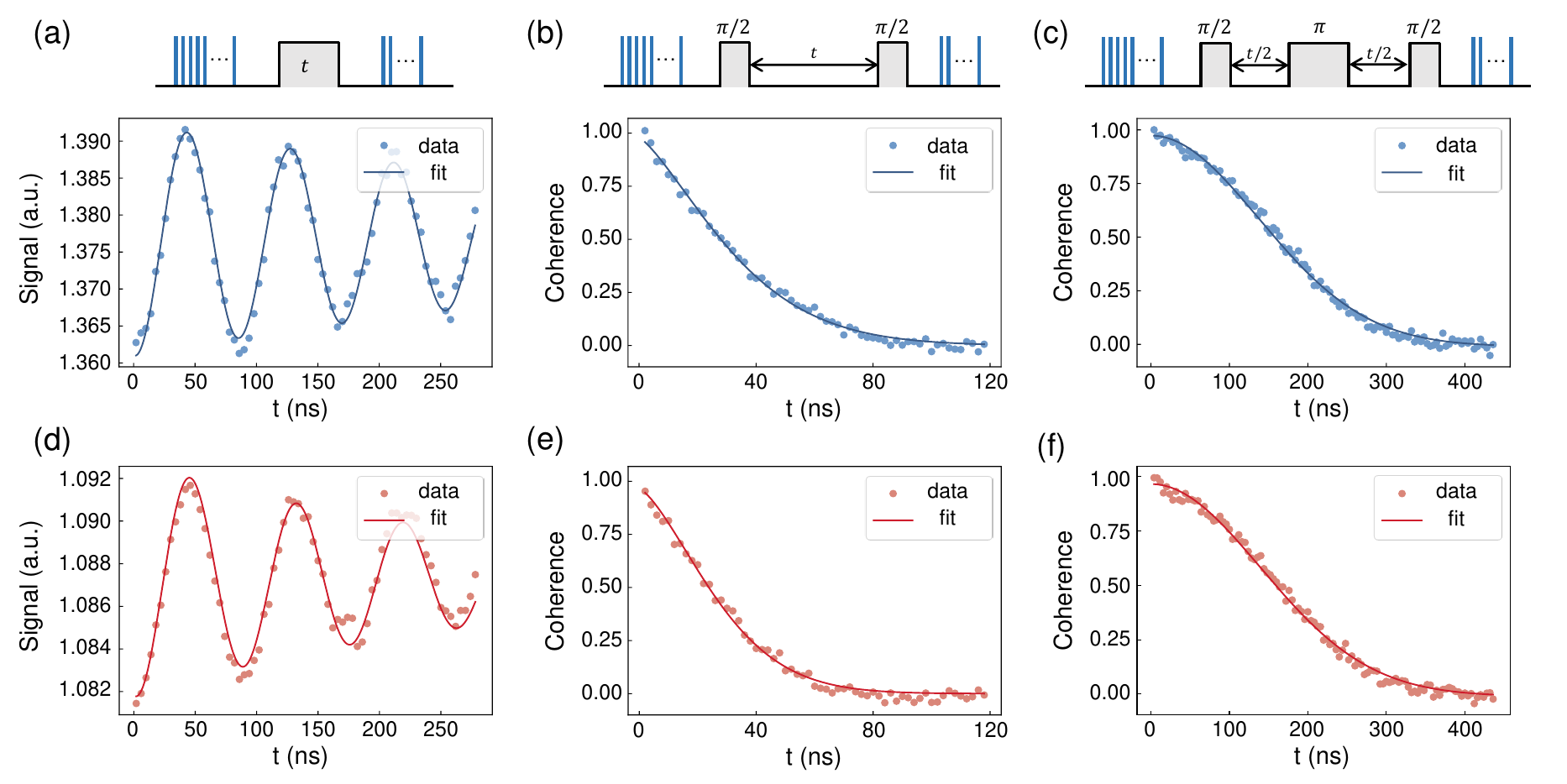}
\caption{
Coherence properties of cerium electron spins. The number of femtosecond laser pulses for spin initialization and readout are 500 and 50, respectively.
(a) Rabi oscillations of cerium electron spins in the film. The signal is fitted with the exponentially decaying cosine function.
(b) FID decay of the cerium electron spins in the film. The decoherence time is fitted as $T_2^*=39\ \rm{ns}$.
(c) Hahn echo decay of the cerium electron spins in the film. The decoherence time $T_2$ is fitted with a Gaussian function, where $T_2=194\ \rm{ns}$.
(d) Rabi oscillations of cerium electron spins in the bulk crystal. 
(e) FID decay of the cerium electron spins in the bulk crystal. The decoherence time is fitted as $T_2^*=33\ \rm{ns}$.
(f) Hahn echo decay of the cerium electron spins in the bulk crystal. The decoherence time $T_2$ is fitted with a Gaussian function, where $T_2=197\ \rm{ns}$.
}\label{fig4}
\end{figure*}

A photoluminesecence image of the film is recorded with the confocal microscope at room temperature, as shown in Fig.~\ref{fig2}(b). The emission spectrum of the bright area shows broad sidebands due to the strong phonon coupling, as shown in Fig.~\ref{fig2}(c), which is in accordance with the results of single cerium ion \cite{kolesov_mapping_2013}. The sample is then cooled down to cryogenic temperature, and the spectrum shows the characteristic ZPL of cerium ions accompanied by the strong phonon sidebands (Fig.~\ref{fig2}(d)). The spectrum is similar to the results of single cerium ion \cite{siyushev_coherent_2014}, which confirms the existence of cerium ions. Furthermore, the film is photo-stable and the photon count rate is above 1 giga counts per second (Gcps), much higher than that of a single cerium ion. The results indicate that cerium ions are uniformly distributed in the film, and the charge states of the ions are stable. 

In order to investigate the spin coherence properties of cerium ions, we perform ODMR experiments at cryogentic temperature. The detailed energy levels of cerium ions in YAG bulk crystals are depicted in Fig.~\ref{fig3}(a), where the spin states correspond to the ground $\rm 4f(1)$ doublet and excited $\rm 5d(1)$ doublet. Under magnetic fields, the electron spin levels split into $\{\rket{{\uparrow}},\rket{{\downarrow}}\}$. The effective ground state Hamiltonian is,
\begin{equation}
\boldsymbol{H}_{\mathrm{ion}}^{\mathrm{eff}}=\mu_{\mathrm{B}} \vec{B}_0 \cdot \boldsymbol{g}_{\mathrm{eff}} \cdot \vec{\sigma},
\label{eq:hamiltonian}
\end{equation}
where the g-tensor is $\boldsymbol{g}_{\mathrm{eff}}=[g_x, g_y, g_z]=[1.87, 0.91, 2.74]$ \cite{lewis_paramagnetic_1966}, exhibiting anisotropy. Cerium ions can occupy six inequivalent sites in the YAG lattice, as shown in Fig.~\ref{fig3}(c). Each site is represented by a cuboid, following the conventions in \cite{van_der_ziel_optical_1971}. The longest, intermediate, and shortest dimensions of the cuboids corresponds to the x, y and z principal axis of $\boldsymbol{g}_{\mathrm{eff}}$, respectively. In the experiment, a magnetic field of $\rm 310\ Gs$ is applied to lift the degeneracy of the energy levels. The direction of magnetic field $\vec{B_0}$ is aligned parallel to the laser beam, and perpendicular to the film surface. Consequently, one of the principal axes of site-1 and site-2 is aligned with the magnetic field, while sites 3 through 6 are magnetically equivalent.

The first step to manipulate cerium electron spins is spin state initialization. Under the excitation of circularly polarized laser $\sigma_+$, the spin-flip transition from $\{\rm 4f(1),{\downarrow}\}$ to $\{\rm 5d(1),{\uparrow}\}$ is stronger than the other three transitions for more than two orders of magnitude \cite{kolesov_mapping_2013}, while the fluorescence probability from $\rm 5d(1)$ doublets to $\rm 4f(1)$ doublets is nearly equal. As a result, spins can be optically initialized to $\rket{{\uparrow}}$ with circularly polarized laser $\sigma_+$, leading to a theoretical spin polarization of 99.7\%. Here $\rket{{\uparrow}}$ ($\rket{{\downarrow}}$) corresponds to the dark (bright) state in the ODMR experiments, respectively. To characterize the spin initialization process experimentally, we excite the ions with a train of circularly polarized laser pulses, and record the fluorescence intensity, as shown in Fig.~\ref{fig3}(b). The fluorescence intensity is at the maximum when starting at the thermal state. After 500 laser pulses, the intensity drops, which corresponds to the polarization process of the cerium spins. The average polarization of the cerium spins and the optical pumping efficiency thus are estimated to be 11.5\%. The relatively low polarization of the cerium spins may be attributed to the higher experiment temperature of the cryostat, and the non-perfect polarization of the laser photons. As a result of the initialization dynamics, spin states can be optically read out by collecting fluorescence photons within an appropriately selected detection window.

The initialized electron spins can be manipulated using resonant microwave fields. To determine the resonant frequency, continuous-wave ODMR experiments are conducted. In the experiment, the frequency of a microwave source is swept while the electron spins are continuously excited by the laser pulses. When the microwave frequency is on resonance, the electron spins are flipped, which is indicated by the increase of the fluorescence intensities. The ODMR results, dipicted in Fig.~\ref{fig3}(d), show resonant frequencies at 397 MHz, 809 MHz and a cluster of four near-degenerate frequencies around 950 MHz. The three peaks correspond to the cerium ions at site-2, site-1 and the other sites, respectively, as marked in Fig.~\ref{fig3}(c). The resonant frequencies match well with the theoretical results calculated with the g-tensor $\boldsymbol{g}_{\mathrm{eff}}$ \cite{SM}. The results also confirms that the (110) crystallographic orientation of the original bulk crystal is preserved in the film.

In the subsequent experiments, we concentrate on the cerium ensemble at site-2. The corresponding Hamiltonian can be rewritten as
\begin{equation}
\boldsymbol{H}_{\mathrm{ion, 2}}^{\mathrm{eff}}=g_y \mu_{\mathrm{B}} B_0 \sigma_z,
\label{eq:hamiltonian2}
\end{equation}
To characterize the coherence properties of the cerium spins, it is necessary to coherently manipulate the cerium spins and detect Rabi oscillation between the states $\{\rket{{\uparrow}}, \rket{{\downarrow}}\}$. The circularly polarized laser pulses are chopped in time sequence with the pulse generator, where 500 laser pulses are used to initialize the electron spin to the polarized state, according to the initialization dynamics in Fig.~\ref{fig3}(b). Sequentially, a resonant microwave pulse with a variable duration is applied to coherently drive the electron spins at ground state. Finally, the spin state is read out with 50 laser pulses. The pulse sequence and corresponding results are presented in Fig.~\ref{fig4}(a). 

The coherence properties of cerium spins are measured with free inductive decay (FID) and Hahn echo sequences, as shown in Fig.~\ref{fig4}(b, c). The measured values of the coherence times are $T_2^*=39\ \mathrm{ns}$ and $T_2=194 \ \mathrm{ns}$, respectively. The coherence times are shorter than those previously reported for single cerium ions \cite{siyushev_coherent_2014}, which may be attributed to the denser cerium electron spin bath, and lattice damages introduced by the film fabrication processes. The implantation of high-dose and high-energy ions creates defects in the lattice and potentially deteriorate the spin coherence properties of rare-earth ions. To evaluate the impact, additional pulse experiments are performed on a Ce:YAG bulk sample with the same experimental setup as a contrast. The sample is the original bulk crystal used to fabricate the film, ensuring that the spin bath environment remains identical to that of the film. The results, presented in Fig.~\ref{fig4}(d-f), show coherence time of $T_2^*=33\ \mathrm{ns}$ and $T_2=197\ \mathrm{ns}$, demonstrating no significant difference from the film. In addition to FID and Hahn echo, the coherence time of cerium spins can be further extended by utilizing dynamical decoupling sequences, such as Waugh-Huber-Haeberlen (WAHUHA) and Carr–Purcell–Meiboom–Gill (CPMG) sequences \cite{waugh_approach_1968, carr_effects_1954, meiboom_modified_1958}. The coherence times achieved through the dynamical decoupling sequences are also comparable to the results of single cerium spins \cite{SM}. These results demonstrate that the spin coherence properties of the cerium ions are effectively maintained through the fabrication processes.

\section{Conclusion}
We have fabricated large-scale, transferrable YAG films with the "smart-cut" technique. The films exhibit micrometer-level thickness and sub-nanometer surface roughness, which are crucial for the integration with micro-cavities. Notably, the spin coherence properties of the rare-earth ions embedded within the films were investigated, exhibiting coherence times comparable to those observed in the bulk crystals. The findings demonstrate the compatibility between the bulk-like spin properties and the thin-film fabrication technique, which is crucial for the development of scalable nanophotonic quantum devices. 
Based on these results, the films can be integrated into Fabry-P{\'e}rot cavity to reduce the mode volume and enhance the Purcell effect. Additionally, the films facilitate effective integration with high-reflectivity materials and support the fabrication of on-chip cavities, such as bullseye cavities \cite{hekmati_bullseye_2023, SM}. These cavities can improve photon collection efficiency and enable the detection and control of single rare-earth ions, advancing researches in quantum information science. Beyond the research of single quantum systems, the films can be integrated into hybrid quantum structures. This integration enables the combined exploitation of optical and spin properties across different quantum systems, thus harnessing the ultra-narrow optical homogeneous linewidths of rare-earth ions and the excellent spin properties of other color centers, such as nitrogen-vacancy center in diamond, presenting potential applications in quantum sensing and quantum communication \cite{chia_hybrid_2024}.

Based on the ion-implantation technique, the thickness of the films can be further reduced by optimizing the implantation energy or subsequent etching processes, which enables a smaller mode volume and enhances the performance of the micro-cavities. Larger films can be fabricated by increasing the spacing between the grooves. The sizes of the films are essentially constrained by the sizes of the bulk crystal and the ion beam, which are on the scale of tens of millimeters. Furthermore, this method is applicable to a wide range of host crystal materials, such as yttrium orthosillicate \cite{SM}, expanding its applications across various rare-earth ion platforms. This versatility opens up new opportunities for device engineering in quantum technologies while preserving the exceptional properties of rare-earth ions.

\

\textit{Supporting information.---} The supporting information contains the schematic diagram of experimental setup, the thickness and roughness of the films, the Zeeman splittings of cerium spins at different sites, the coherence properties of cerium spins with dynamical decoupling sequences, the methods to improve collection efficiency, and the simulation of applying smart-cut technique to yttrium orthosilicate (YSO) crystals.

\textit{Competing interests.---} The authors declare no competing interests.

\textit{Data availability. ---} All data needed to evaluate the conclusions are present in the paper.

\begin{acknowledgments}
\textit{Acknowledgements.---} This work is supported by the National Natural Science Foundation of China (Grants No.\ 12274400, 12122508, 92265204, T2325023, 123B1019),  the Innovation Program for Quantum Science and Technology (Grants No.\ 2021ZD0302200), the China Postdoctoral Science Foundation (Grants No.\ 2024M763128), the Postdoctoral Fellowship Program of CPSF (Grants No.\ GZC20241652) and the Fundamental Research Funds for the Central Universities.
\end{acknowledgments}

\textit{Notice. ---} This document is the unedited author’s version of a submitted work that was subsequently accepted for publication in ACS Photonics (\textcopyright American Chemical Society) after peer review. To access the final edited and published work, see \href{https://doi.org/10.1021/acsphotonics.4c02520}{https://doi.org/10.1021/acsphotonics.4c02520}


%

\end{document}


\title{Supporting Information for "Coherence properties of rare-earth spins in micrometer-thin films"}
	\maketitle
	
	\renewcommand{\thefigure}{S\arabic{figure}}
	\setcounter{figure}{0}

\begin{figure}[h]
	\includegraphics[width=0.9\columnwidth]{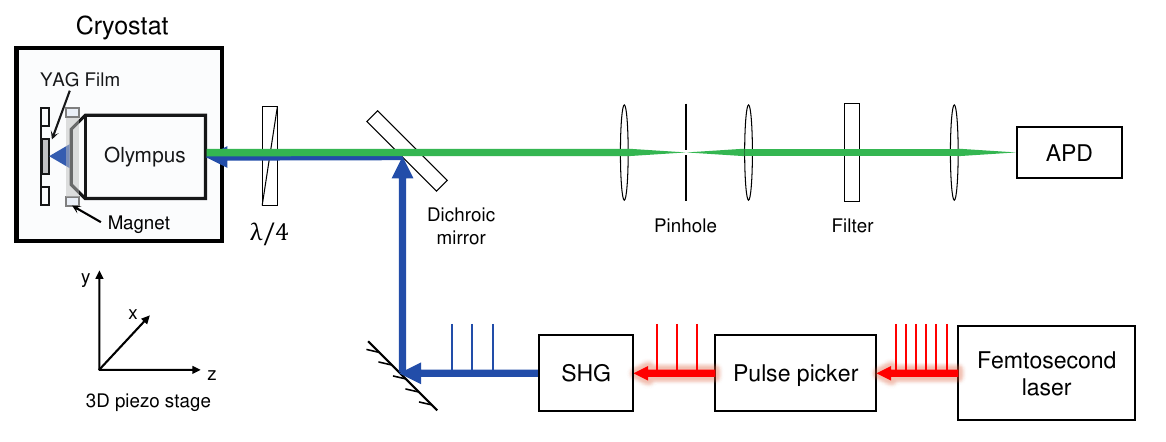}
	\caption{Experimental setup used to locate the film, excite the cerium ions, and study the properties of cerium electron spins. A ring-shaped magnet is fixed on the Olympus objective lens to apply the magnetic field $\vec{B}_0$ on yttrium aluminum garnet (YAG) films.
	}\label{figs0}
\end{figure}

\section{The thickness and roughness of the films}
\begin{figure}[h]
	\includegraphics[width=0.9\columnwidth]{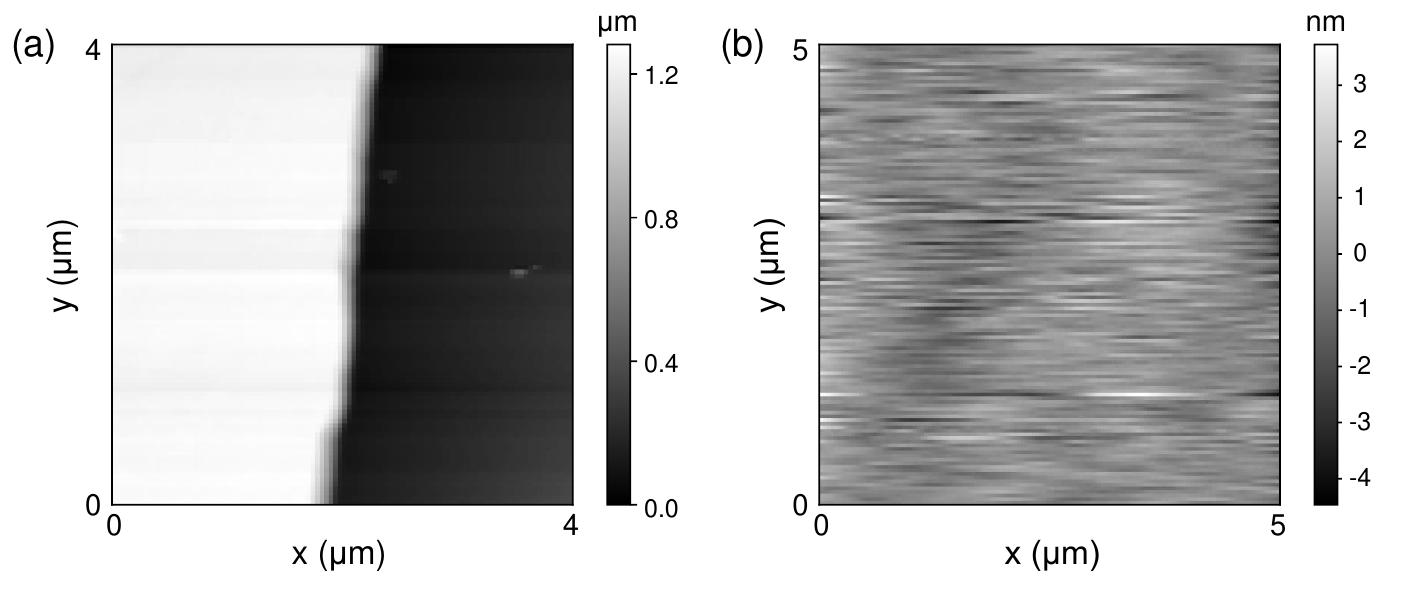}
	\caption{Atomic force microscope (AFM) characterization of the films. (a) The thickness of the film is determined to be 1.22 $\rm \mu m$. (b) The surface roughness of the film is measured at 0.717 $\rm nm$ rms over a 5 $\rm \mu m$ scan area.
	}\label{figs1}
\end{figure}

\begin{figure}[h]
	\includegraphics[width=0.9\columnwidth]{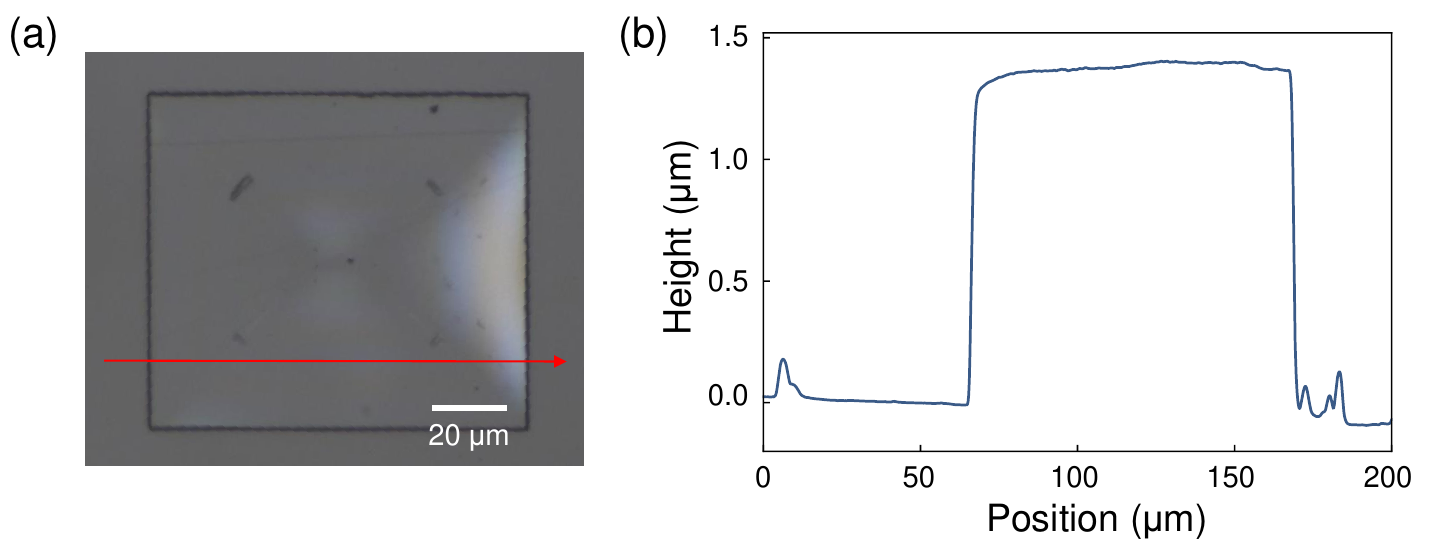}
	\caption{(a) Microscope image of a transferred YAG film on sapphire substrate. The red arrow marks the trajectory measured in (b). (b) Height profile of the film sample measured with Bruker Dektak XT.}\label{figs1-1}
\end{figure}

The thickness and roughness of the YAG films are characterized by atomic force microscope (AFM), as shown in Fig.~\ref{figs1}. The film shown in Fig.2 of the main text is not bonded well, mainly due to surface protrusions on the sapphire substrate. Another YAG film is bonded to a new substrate, as illustrated in Fig.~\ref{figs1-1}. The height profile of this film sample is measured with Bruker Dektak XT. The root-mean-square (rms) height variation of the sample over a scanning range of 85 $\rm \mu m$ is measured to be 14.4 nm. 


\section{Zeeman splittings of cerium spins at different sites}

\begin{figure}[ht]
	\includegraphics[width=0.6\columnwidth]{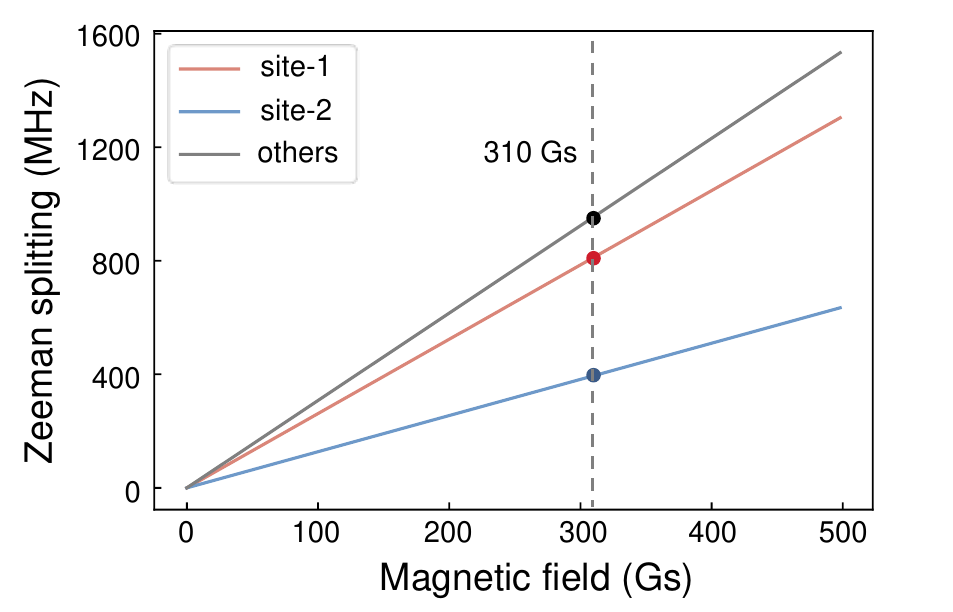}
	\caption{Zeeman splitting of the cerium electron spins for the magnetic field parallel to the [110] crystallographic direction. The solid lines represent the simulated results, and the points represent experimental results at 310 Gs.}\label{figs2}
\end{figure}

Cerium ions can occupy six inequivalent sites in YAG crystal. Because of the anisotropy of g-factor, Zeeman splitting of cerium electron spins depends on both the magnitude and direction of the external magnetic field $\vec{B}_0$. Generally, this would result in six resonances in the continuous-wave ODMR spectrum. However, the number of resonances would reduce for certain directions, due to the symmetry of YAG lattice. In our experiment, the film is fabricated from a [110]-oriented YAG crystal, and the magnetic field is applied perpendicular to the surface of the film. As a result, the magnetic field $\vec{B}_0$ is aligned with the $x$-axis ($y$-axis) of site-1 (2), and the other sites are magnetically equivalent. The experimental results in the main text and the simulation results are shown in Fig.~\ref{figs2}, where the deviations might result from the deviation of the crystal field parameters and misalignment of the magnetic field.

\section{Coherence properties of cerium spins with dynamical decoupling sequences}

\begin{figure}[h]
\includegraphics[width=0.95\columnwidth]{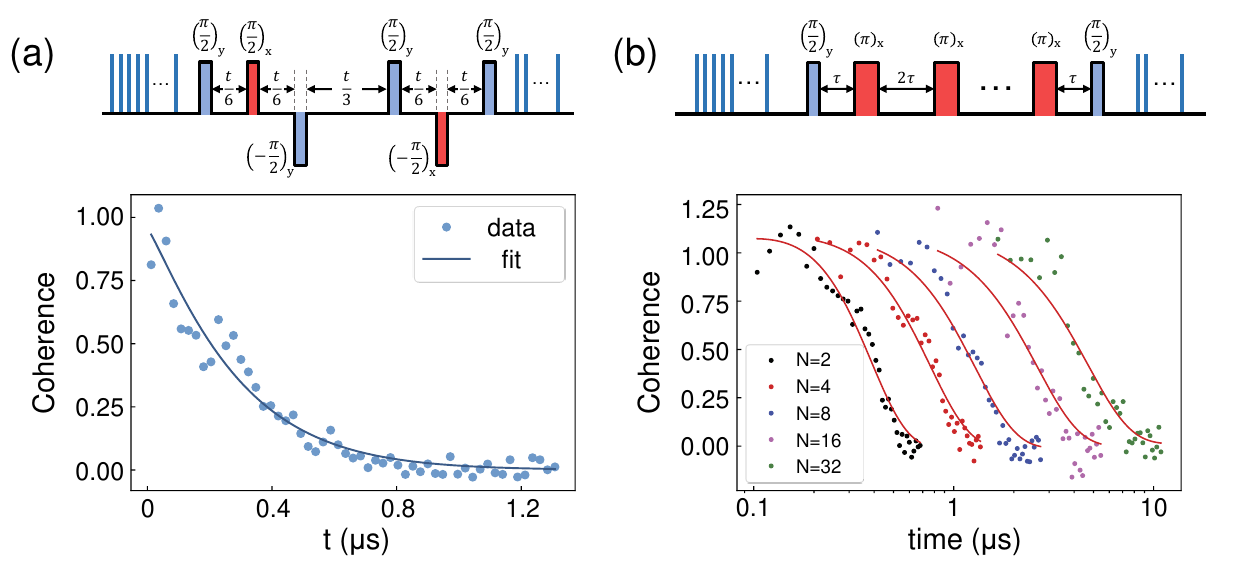}
\caption{Dynamical decoupling of cerium spins from spin bath. The number of femtosecond laser pulses for spin initialization and
readout are 500 and 50, respectively. (a) WAHUHA sequence and the experimental results of the cerium spins in the film. The decoherence time is fitted as $T_2=297 \rm \ ns$. (b) CPMG sequence and the experimental results of the cerium spins for several sequences. N represents the number of $\pi$ pulses in the corresponding sequence. The data are fitted with the red curves. }\label{figs2-1}
\end{figure}

In addition to the FID and Hahn echo sequences discussed in the main text, the coherence time of cerium electron spins can be further increased with dynamical decoupling sequences, such as Waugh-Huber-Haeberlen (WAHUHA) and Carr–Purcell–Meiboom–Gill (CPMG) sequences \cite{waugh_approach_1968, carr_effects_1954, meiboom_modified_1958}. The WAHUHA sequence, which utilizes $\pi/2$ pulses, (Fig.~\ref{figs2-1}(a)), involves rotation of the spin operator that effectively cancel the dipolar couplings. In the experiments, the coherence time is extended to $T_2=297 \rm \ ns$ with WAHUHA sequence. The CPMG sequences consist of equidistant $\pi$ pulses (Fig.~\ref{figs2-1}(b)), where the number of $\pi$ pulses is N. The coherence time achieved with the CPMG sequences is comparable to the results of single cerium ions in bulk crystal, with an equivalent number of pulses \cite{siyushev_coherent_2014}.

\section{Methods to improve collection efficiency}
\begin{figure}[ht]
\includegraphics[width=0.8\columnwidth]{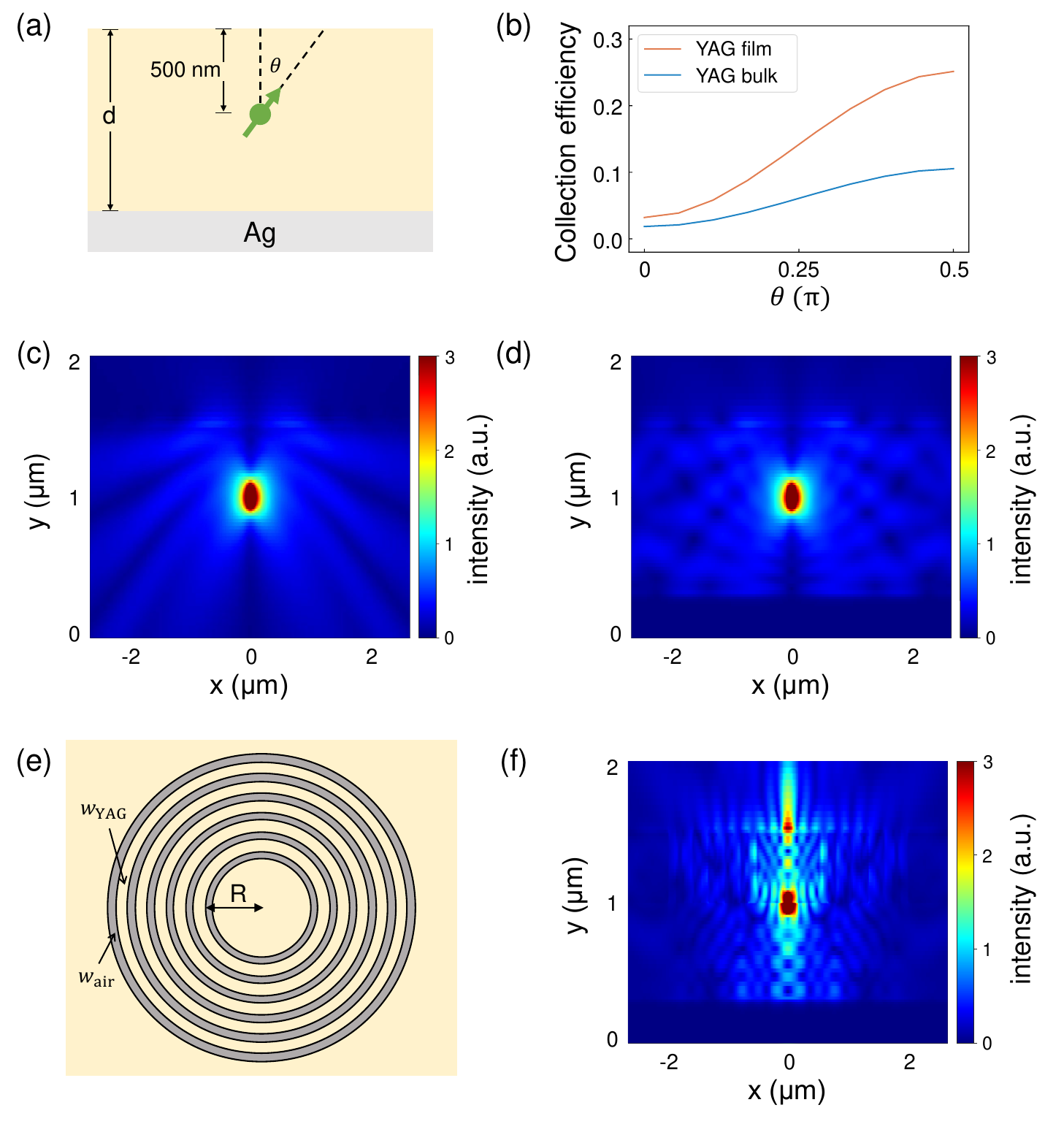}
\caption{
(a) Schematic diagram of films on a reflective layer.
(b) Collection efficiencies of an electric dipole in YAG bulk and films on a reflective layer, respectively.
(c) Distribution of the electric field in bulk crystal. The simulation is performed with $\theta=0$ and the thickness of the bulk crystal is set as $\rm {d=500\ \mu m}$. The resulting collection efficiency is 1.85\%.
(d) Distribution of the electric field in films with an Ag reflective layer. The thickness of the film and the Ag layer is $\rm{d=1.2\ \mu m}$ and 100 nm, respectively. The resulting collection efficiency is 3.2\%.
(e) Schematic diagram of the bullseye structure. 
(f) Distribution of the electric field in the thin films with bullseye structure in (e). The simulation is performed with $\theta=0$, $R=500\ \rm{nm}$, $w_{\rm YAG}=120\ {\rm nm}$, and $w_{\rm air}=130\ {\rm nm}$. The etched depth of the bullseye structure is 500 nm.}\label{figs3}
\end{figure}

Thin films can be integrated on high-reflectivity materials to enhance the collection efficiency. For example, reflective layers can be deposited on the films to reflect the fluorescence, thereby enabling more fluorescence to be collected by the objective lens (Fig.~\ref{figs3}(a)). We use Finite Difference Time Domain (FDTD) methods to compare the emission of an electric dipole in YAG bulks and films with 100 nm-thick reflective Ag layer, and simulate the fluorescence collection efficiency. The depth of the dipole is set at 500 nm, and the thickness of the film is set at 1.2 $\mu \rm{m}$. Collection efficiency is determined by calculating the ratio of the far-field power collected by the objective lens to the total emission power of the dipole. As shown in Fig.~\ref{figs3}(b), collection efficiency is improved effectively across the angular range of $0-\pi/2$, which confirms the effectiveness of integrating thin films onto reflectivity materials. The distribution of the emitted electric fields in bulk and film is shown in Fig.~\ref{figs3}(c) and (d), respectively, where the angle $\theta$ is set at $0$. Furthermore, collection efficiency can be enhanced by fabricating bullseye structures on the thin films (Fig.~\ref{figs3}(e)), which modify the emission direction of the dipole at the surface, and effectively reduce the refractive index of the surface to suppress internal reflection. In the simulations, the estimated collection efficiency exceeds 60\% (Fig.~\ref{figs3}(f)), which exhibits the potential to efficiently detect and control single rare-earth ions with microstructures on thin films.

\section{Smart-cut towards yttrium orthosilicate crystals}

\begin{figure}[ht]
\includegraphics[width=0.8\columnwidth]{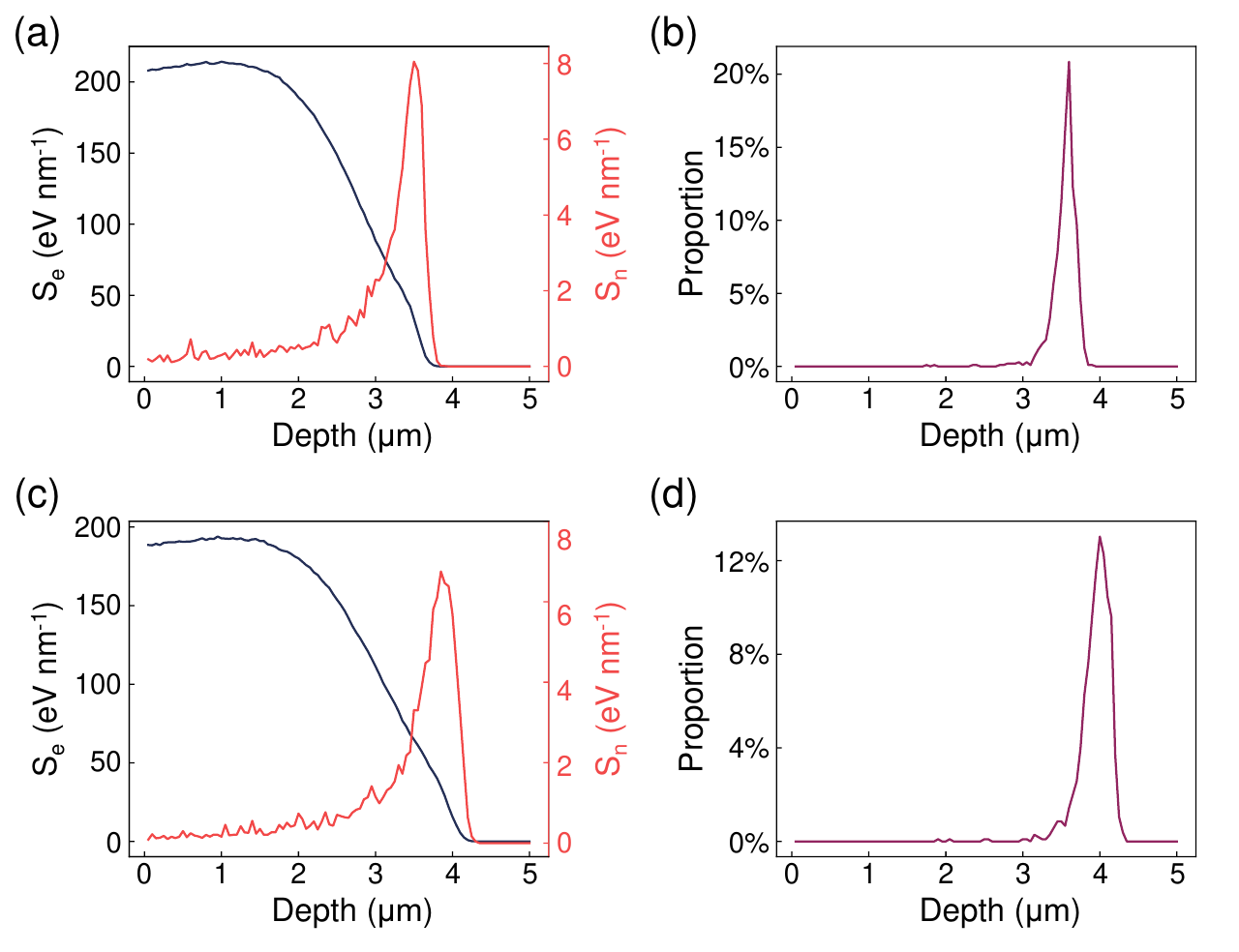}
\caption{
Simulation of the ion implantation processes in YAG and YSO by SRIM-2013. (a) The electronic energy loss ($S_e$) and nuclear energy loss ($S_n$) of carbon ions versus the injection depth in YAG. (b) The depth distribution of the injected carbon ions in YAG. (c) The electronic energy loss ($S_e$) and nuclear energy loss ($S_n$) of carbon ions versus the injection depth in YSO.  (d) The depth distribution of the injected carbon ions in YSO.
}\label{figs4}
\end{figure}

Yttrium orthosilicate (YSO) is a widely used materials serving as host of rare-earth ions. The application of smart-cut techniques to YSO crystals will facilitate the research across various rare-earth ion platforms, such as Er:YSO, Eu:YSO and Nd:YSO. We utilized SRIM-2013 to simulate the results of implanting high-energy (6 MeV) carbon ions into YAG and YSO crystals. During implantation, incident carbon ions interact with the electrons and nucleus in the crystals through inelastic and elastic scattering, resulting in electronic energy loss $S_e$ and nuclear energy loss $S_n$. Fig.~\ref{figs4} (a) and (c) demonstrates the electronic (nuclear) energy loss versus implantation depth for YAG and YSO crystals, respectively. Consequently, the carbon ions stop inside the crystals, as shown by the stopping range distributions in Fig.~\ref{figs4} (b) and (d) for YAG and YSO. This process creates a damaged layer beneath the crystal surface, which can be corroded by acid. As a result, thin films of YSO can be exfoliated from the bulk crystals.

%